\documentstyle[prl,aps,multicol,epsf]{revtex}
\def\lsim{\:\raisebox{-0.5ex}{$\stackrel{\textstyle<}{\sim}$}\:}
\begin{document}
\draft
\title{Travelling waves in a drifting flux lattice}
\author{R.Aditi Simha\cite{a} and Sriram Ramaswamy\cite{b,c}}  
\address{Department of Physics, Indian
  Institute of Science, Bangalore 560 012, India}
\date{final version published in Phys. Rev. Lett. {\bf 83} (1999) 3285 }
\maketitle
\begin{abstract}
Starting from
the time-dependent Ginzburg-Landau (TDGL) equations for
a type II superconductor, we derive the equations of motion
for the displacement field of a moving vortex lattice ignoring
pinning and inertia. We show that it is linearly stable and, surprisingly, 
that it supports {\em wavelike} long-wavelength
excitations arising {\em not} from inertia or elasticity but from the
strain-dependent mobility of the moving lattice. It should be possible
to image these waves, whose speeds are a few $\mu$m/s, using fast scanning
tunnelling microscopy.
\end{abstract}
\begin{multicols}{2}
It was shown in \cite{sr}, on general symmetry grounds, 
that an ordered array of particles moving through a dissipative medium   
({\it e.g.}, a steadily sedimenting colloidal crystal or a  
flux-point lattice drifting through a type II superconductor) is 
governed by dynamical equations qualitatively different 
from those for a lattice at thermal equilibrium. 
Even the long-wavelength dynamical stability of such a drifting lattice 
was shown to rest not on its elasticity but 
on the signs of certain phenomenological parameters 
[see eqns. (\ref{xeqn}) and (\ref{yeqn}) below] 
governing the dependence of the local mobility on the lattice strain. 
A microscopic calculation \cite{crow} showed (see \cite{sr}) 
that for a sedimenting colloidal crystal the signs were such  
as to lead to an {\em instability}. We know of no analogous calculation 
for driven flux lattices.  
In this Letter we ask: are drifting flux lattices stable? We answer 
this question of  fundamental importance starting from a time-dependent 
Ginzburg-Landau (TDGL) treatment {\em without quenched disorder}. We find that 
the moving lattice state is {\em stable}, with small-amplitude, 
long-wavelength disturbances propagating as {\em underdamped 
waves} whose speed, we emphasise, is determined by the {\em strain
-dependent mobility} and the imposed current, and {\em not} by inertia 
and flux-lattice elasticity. We calculate the wavespeed (a few 
$\mu$m/s) in terms of independently measurable parameters arising 
in the TDGL equations. 

We begin by summarising the derivation of the coarse-grained 
dynamical equations for a drifting lattice \cite{sr} and  
defining the quantities 
we are going to calculate. Consider a slab of type II superconductor 
of thickness much larger than the magnetic penetration depth $\lambda_H$,  
lying in the $xy$ plane, threaded by a flux lattice (spacing $\lsim 
\lambda_H$) with magnetic field along the $z$ direction. 
An applied spatially uniform transport 
current density ${\bf J}_t = J_t \hat{\bf x}$, gives a Lorentz force 
$- J_t \phi \hat{\bf y}/c$ per unit length on a vortex carrying flux 
$\phi \hat{\bf z}$, $c$ being the speed of light.  
The perfect flux-point lattice will then acquire a constant, 
spatially uniform drift speed $v_L = M J_t \phi /c$. Here 
$M$, the macroscopic mobility of the lattice, is determined 
by dissipative processes in the normal core as well as by the relaxation 
of the electromagnetic and order-parameter fields in the region 
between the vortices. Any perturbation of the perfect moving lattice 
will result in inhomogeneities in the local 
electromagnetic and order parameter fields, and thus to a spatially 
varying flux-point velocity. The mobility is thus  
a tensor which  
depends on the local state of distortion of the flux lattice. 
For a lattice drifting along $- \hat{\bf y}$, ignoring Hall effects, 
pinning, inertia \cite{dmgtvr}, and the effects of lattice periodicity,  
the displacement field ${\bf u}=(u_x,u_y)$ as a function of position 
${\bf r}$ and time $t$, defined with respect to a perfectly ordered 
crystal, in a frame co-moving on average with the flux lattice,  
must then obey \cite{sr}  
\begin{eqnarray}
\label{xeqn}
\partial_t u_x = v_1\partial_y u_x + v_2\partial_x u_y+
    D_T \nabla^2 u_x +D_L \partial_x ^2 u_x + \nonumber \\
D_L \partial_x\partial_y u_y     +   O(\nabla u \nabla u) \mbox{;} 
\end{eqnarray}
\begin{eqnarray} 
\label{yeqn} 
\partial_t u_y = v_3\partial_x u_x+v_4 \partial_y u_y +
   D_T \nabla^2 u_y +D_L \partial_y ^2 u_y + \nonumber \\
D_L \partial_x\partial_y u_x    + O (\nabla u \nabla u) \mbox{,}
\end{eqnarray}
where the terms containing the phenomenological coefficients \cite{ftn1}  
$v_i \propto v_L$ arise from the ``hydrodynamic'' interaction 
of the moving vortices, $D_L= M( \lambda+ 2\mu )$, 
and $D_T = M \mu $, $\lambda$ and $\mu$ being the Lam\'e coefficients 
of the flux lattice \cite{ftn2}. These equations are constructed using 
general symmetry arguments and hold for any steadily drifting lattice 
at large length scales ($\gg \lambda_H$, for a flux lattice).
In this Letter we calculate the coefficients $v_i$ for the 
specific case of a drifting flux lattice, 
from a time-dependent Ginzburg-Landau (TDGL) description
to which we turn next. The importance of the $\{v_i\}$ for 
the long-wavelength behaviour of the drifting flux lattice is 
clear: $v_2 v_3 > 0$ yields a wavelike dispersion whereas $v_2 v_3 < 0$  
a linear instability.  

Scaling lengths by $\lambda_H$, 
energies by the condensation energy $E_c$ in a volume $\lambda_H^3$, 
the order parameter by its bulk mean-field value in the 
superconducting phase, times by $\hbar/E_c$, the magnetic 
field ${\bf H}$ by $\sqrt{2} H_c$ where $H_c$ is the thermodynamic 
critical field, the total electrochemical  
potential by $E_c /e^*$ where 
$e^* = 2 e$ is the charge of the Cooper pair, and defining 
the Ginzburg-Landau parameter $\kappa = \lambda_H/\xi$ where 
$\xi$ is the bare coherence length, we obtain the  
dimensionless TDGL equations \cite{tdgl,schmid,dor}  
for the dynamics of the superconducting order 
parameter $\psi({\bf r},t$):  
\begin{equation}
\label{tdgl} 
  (\partial _t + i\Phi)\psi=\Gamma \left[\left(\frac{\nabla}{\kappa}-i{\bf A}
  \right)^2  \psi +\psi -\mid \psi\mid ^{2} \psi \right]\mbox{ ,}
  \end{equation} 
where the phenomenological kinetic coefficient $\Gamma$ is in 
general complex, with real and imaginary parts $\Gamma_1$ and $\Gamma_2$ 
respectively. 

 The equation of motion for the vector potential is given
by Amp\`{e}re's law,
   \begin{equation}
\label{ampere}
  \nabla \times\nabla \times{\bf A}={\bf J}_n+{\bf J}_s \mbox{ ,}
\end{equation}
  where the normal and super currents are, respectively,   
\begin{eqnarray}
\label{currents}
  {\bf J}_n= {\bf \sigma}.\left[-\frac{{\bf \nabla} \Phi}{\kappa}-\partial _t
  {\bf A}\right] \mbox{ ,} \nonumber \\
  {\bf J}_s=\frac{1}{2\kappa i}(\psi^*{\bf \nabla}\psi-\psi{\bf \nabla}\psi^*)
  -\mid \psi\mid^2{\bf A} \mbox{ ,}
\end{eqnarray}
${\bf \sigma}$ being the normal-state conductivity tensor.  
 
We work in the large $\kappa $ limit, where a  
phase-only approximation of the TDGL equations (\ref{tdgl}) applies,   
and, for simplicity, we set the normal state Hall conductivity  
$\sigma_{xy} = 0$.
We begin by writing $\psi$ in terms of an amplitude $f$ and a phase 
$\chi$:  
\begin{equation}
\label{aph}
\psi({\bf r},t)=f({\bf r},t)\mbox{exp}[i\chi({\bf r},t)].
\end{equation} 
In terms of the gauge-invariant vector and scalar potentials,
${\bf Q =A-\nabla}\chi/\kappa$ and $P=\Phi+\partial _t\chi$,
the magnetic and electric fields are then, respectively, 
\begin{equation}
\label{fields}
{\bf h}={\bf \nabla} \times {\bf Q} \mbox{ ,} \hspace{.7cm}
{\bf E}= -\frac{{\bf \nabla} P}{\kappa}-\partial _t{\bf Q} \mbox{ .}
\end{equation}

For large $\kappa$, deep in the superconducting phase, the amplitude 
relaxes rapidly to a value determined 
by the phase. We can thus solve for $f$ in terms of 
$\chi$ from (\ref{tdgl}) yielding, to leading order in $1/\kappa$,  
the effective phase-only TDGL equation
\begin{equation}
\label{pheff} 
   \partial_t\chi +\Phi=P=-{\bf \nabla.Q}/\gamma_1 \kappa \mbox{ .}
   \end{equation}  
with 
$\gamma_1 = \mbox{Re}\left[\Gamma ^{-1}\right]$ 
and  
   \begin{equation}
\label{ampereff}
   \nabla\times\nabla\times{\bf Q}={\bf \sigma}.\left[\frac{-\nabla P}{\kappa}-
   \partial_t{\bf Q}\right]-{\bf Q}\mbox{ .}
   \end{equation}
Lastly, charge conservation --- ${\bf \nabla}.({\bf J}_n+{\bf J}_s)= 0$ --  
with (\ref{currents}) and (\ref{fields}) leads to  
\begin{equation}
\label{cont} 
   {\bf \nabla.(\sigma.E) -\nabla.Q}=0  
\end{equation}
which, with (\ref{pheff}) and (\ref{ampereff}) implies  
   \begin{equation}
\label{peff} 
   \sigma_{xx}\frac{{\bf \nabla .}}{\kappa}\left[-\frac{{\bf \nabla}P}{\kappa}-
   \partial_t{\bf Q}\right]
   +\gamma_1P=0 \mbox{ .}
   \end{equation}

The $\{v_i\}$ in (\ref{xeqn}),
(\ref{yeqn}), which encode the change in the mobility of a region of
the flux lattice when it is compressed or tilted,   
arise primarily from electromagnetic field disturbances, 
screened on the scale  
$\lambda_H$ \cite{ftn3}. 
Ideally, therefore, we should calculate the mobility of distorted regions
on a scale $\lambda_H$. However, our main concern is   
the {\em signs} of the $\{v_i\}$, i.e., in the {\em direction}
of drift of a tilted region and in whether a denser region
drifts {\em faster} or {\em slower} than a rarer region.  
To this end, we take the simplest   
compressions/rarefactions and tilts, namely, those taking place 
at the level of a pair of particles. This should give a qualitatively 
correct assessment of the stability and a reasonable  
estimate of the wavespeed. Indeed, our calculation shows 
that the $v_i$s decrease by a factor of 10 as the flux-lattice 
spacing varies from .25 $\lambda_H$ to $\lambda_H$, 
justifying {\it post facto} this nearest neighbour approximation.   
We work, therefore, with a {\em pair} of flux points moving rigidly 
with a velocity ${\bf v}_L$,
as a function of their fixed separation vector $\bf a$.
For such rigid motion, time-derivatives can be replaced by
$-{\bf v}_L.{\bf \nabla}$. Expanding (\ref{pheff}), (\ref{ampereff}),
and (\ref{peff}) in powers of $v_L$, we obtain at O(1)
the equilibrium, time-independent Ginzburg-Landau equations,
and at O(${\bf v}_L$) a set of linear inhomogenous differential
equations.       

Exploiting \cite{gor,dor} the invariance of the time-{\em independent}
Ginzburg-Landau equations under an arbitrary virtual displacement 
${\bf d}$, the requirement of compatibility 
between  ${\bf v}_L$ and the imposed transport current ${\bf J}_t$ leads,   
for large $\kappa$ and within the phase only approximation, 
to the ``solvability condition'' for the inhomogeneous O$(v_L)$ equations:  
\begin{equation}
\label{solvab}
\frac{1}{\kappa}\int d{\bf S}.({\bf J}_{1s}\chi_d-{\bf J}_d\chi_1)=
   \gamma_1\int(\chi_dP)d{\bf r} \mbox{;}
\end{equation}
the integral on the left-hand side is over the boundary 
of the sample, 
${\bf J}_d \equiv {\bf d.\nabla J}_0$, 
$\chi_d \equiv {\bf d.\nabla} \chi_0$, 
$J_0$ and $\chi_0$ being respectively  
the supercurrent and phase field at {\em equilibrium}, and 
the subscripts 0,1 denoting the O(1) and O(${\bf v}_L$) parts 
respectively of the term in question.
Eq. (\ref{solvab}) will yield the relation between ${\bf J}_t = 
J_t \hat{\bf x}$ and ${\bf v}_L$, {\it i.e.} the vortex equation 
of motion. 
\begin{figure}
 \epsfysize=7cm
\centerline{\epsffile{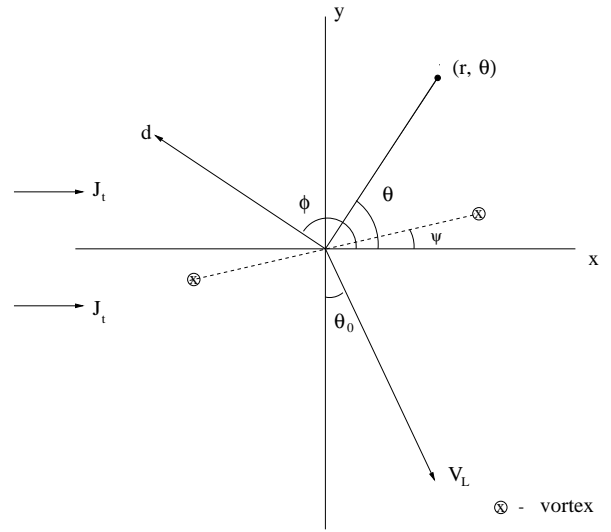}}
\caption{\label{coordsys}\narrowtext{Coordinate system ($r,\theta$) for
two-vortex case}} 
\end{figure}

Consider a pair of identical unit vortices, in a geometry defined in 
Fig.\ref{coordsys} (in cylindrical polar coordinates $(r,\theta,z)$).  
$\psi$ and $\phi$ are the angles made by  
${\bf J}_t$ with ${\bf a}$ and the virtual displacement 
${\bf d}$ respectively, and $\theta_0$ that between ${\bf v}_L$ 
and the negative $ y$-axis. We assume the flux lines to be parallel
to the $z$-axis and ignore the effects of line wandering. 
We also define cylindrical coordinates $ (r_1,\theta_1,z)$
and $(r_2,\theta_2,z)$ with their origins at the two vortices. 
The surface integral on the left-hand side of (\ref{solvab}) 
can be expressed in terms of the applied transport current. 
At the boundaries the fields are effectively those of a single vortex 
at the origin, with twice the winding number.  
Therefore
   ${\bf J}_{1s}(r=\infty ,\theta) ={\bf J}_t $, ${\bf J}_d.{\bf \hat{e}}_r=
   2d \sin(\theta-\phi)/(\kappa r^2) $, $\chi_d=2{\bf d.\nabla}\theta=
   -2d \sin(\theta-\phi)/r$ and $\chi_1=\kappa J_t r \cos \theta$. 
Substituting these expressions into the left-hand side of (\ref{solvab}), 
and performing the angular integration we find
   \begin{equation}
    \frac{1}{\kappa}\int d{\bf S}.[{\bf J}_{1s}\chi_d-{\bf J}_d\chi_1]=
    - 2\frac{2\pi}{\kappa}({\bf J}_t\times{\bf \hat{z}}).{\bf d}\mbox{ .}
    \end{equation}
 Evaluation of the right-hand side of (\ref{solvab}) requires 
solving for $P({\bf r},t)$ from (\ref{peff}) which, at O$(v_L)$, 
is simply  
    \begin{equation}
\label{p1eff}
   \frac{\sigma_{xx}}{\kappa^2} \nabla^2P-\gamma_1P = 0
   \end{equation}
    Near the centre of each vortex
    $P\approx -{\bf v}_L.{\bf \nabla}\chi $ and $\chi$ is equal to the angular
    variable $\theta_{1}$ or $\theta_{2}$ around that vortex. 
Therefore $P\approx v_L 
    \cos(\theta_1-\theta_0)/r_1$ as $r_1\rightarrow 0$ and
    $P\approx v_L \cos(\theta_2-\theta_0)/r_2$ as $r_2\rightarrow 0$.
     The solution to eqn.(\ref{p1eff}) for the vortex pair with these
     boundary conditions is 
   \begin{equation}
\label{p1pair}
   P({\bf r})
   =\tilde{v} [K_1(\alpha r_1) \cos(\theta_1-\theta_0)+K_1(\alpha r_2) 
 \cos(\theta_2-\theta_0)]
\end{equation} 
   where $\tilde{v}= v_L \alpha$ and $\alpha = \kappa \sqrt{\gamma_1
/\sigma_{xx}}$.    Also,    	
   \begin{equation}
\label{chipair} 
   \chi_d={\bf d.\nabla}\chi=d [ \sin(\phi-\theta_1)/r_1+ 
   \sin(\phi-\theta_2)/r_2 ]\mbox{ .}
   \end{equation}
   Using (\ref{p1pair}) and (\ref{chipair}) on the right hand side of 
(\ref{solvab}), and noting that ${\bf d}$ is 
arbitrary, we obtain the vortex-pair equation of motion in the form  
\begin{eqnarray}
\label{2vorteom}
2\frac{2\pi}{\kappa}({\bf J_t\times \hat{z}})=A{\bf v}_L+
B({\bf v}_L\times {\bf \hat{z}})+C({\bf v}_L.{\bf \hat{a}}){\bf \hat{a}}
\nonumber \\
-D{\bf\hat{a}}.({\bf v}_L\times {\bf \hat{z})}{\bf \hat{a}}
   \end{eqnarray}
    where $ A, B, C,$ and $D$ are functions of $\alpha$ and $a = |{\bf a}|$ 
only. All dependence on the angle of tilt $\psi$ is in the scalar 
and vector products in (\ref{2vorteom}). 
Evaluating the integrals, we find that $B=0=D$ (a consequence of the 
phase-only approximation {\em and} the assumption $\sigma_{xy} =0$), and  
$A,C > 0$. 
Inverting (\ref{2vorteom}) we see that  
\begin{equation}
\label{2vortmob}
v_{Li} = M \left[\delta_{ij} - {N \over {1 + N}} 
{a_i a_j \over {a^2} }\right] F_j  
\end{equation}
where ${\bf F} = -{4\pi\over \kappa}J_t\hat{\bf y}$ 
is the Lorentz force, $M = 1/A$, and $N = C/A$. 
(\ref{2vortmob}) differs from 
that for a single vortex \cite{gor,dor} in the $N$ term:  
for $\psi \neq 0$ or $\pi/2$ the centre-of-mass velocity is not 
parallel to the driving force. 

\begin{figure}
\epsfysize=8cm
\epsfxsize=6cm
\centerline{\epsffile{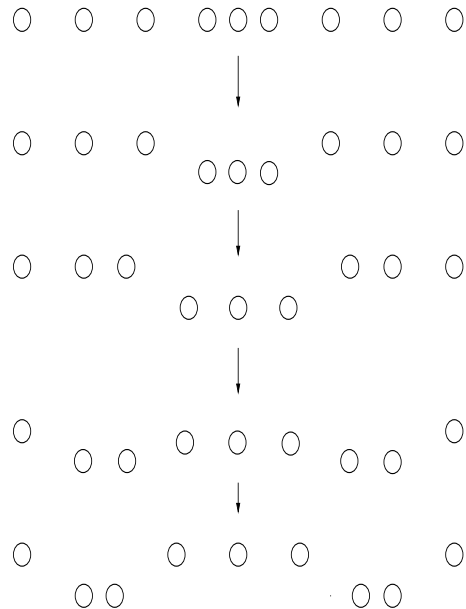}}
\vspace*{1cm}
\caption{\label{fevol}\narrowtext{The wave travelling along $\pm \hat{\bf x}$
that follows  a local compression of an array of vortices moving along
$- \hat{\bf y}$ }}
\end{figure}

Now consider a steadily drifting undistorted flux-point lattice, 
and focus on a nearest neighbour pair of flux points with initial 
separation vector ${\bf a}_0$. Perturb it slightly:  ${\bf a}_0 
\rightarrow {\bf a}_0 + {\bf \delta a}$, thus causing a velocity 
perturbation ${\bf \delta v}$. Then we can extract the $\{v_i\}$ 
by differentiating our two-vortex result (\ref{2vortmob}) as follows: 
if ${\bf a}_0 || \hat{\bf y}$ 
and ${\bf \delta a} || \hat{\bf x}$, then 
${\delta v_x \over {\delta a / a_0}} = v_1$; 
if ${\bf a}_0 || \hat{\bf x}$
and ${\bf \delta a} || \hat{\bf y}$, then
${\delta v_x \over {\delta a / a_0}} = v_2$;
if ${\bf a}_0 || \hat{\bf x}$
and ${\bf \delta a} || \hat{\bf x}$, then
${\delta v_y \over {\delta a / a_0}} = v_3$; 
if ${\bf a}_0 || \hat{\bf y}$
and ${\bf \delta a} || \hat{\bf y}$, then
${\delta v_y \over {\delta a / a_0}} = v_4$; 
We have assumed, as justified early in the paper, that changes in vortex 
velocity due to a local distortion are local.  
We find: (a) $v_2 > 0$, so that a lattice moving in 
the $- \hat{\bf y}$ direction will veer to the right(left) if the horizontal 
crystal planes are tilted up to the right (left); and (b)  
$v_3 > 0$, so that a local $x$-compression of the lattice 
increases the velocity in the direction of the force. 
From (\ref{xeqn}) and (\ref{yeqn}), 
this means the moving flux-lattice is {\em stable}. Also, $v_1 > 0 $ 
and $v_4 < 0$ for the coefficients controlling the wavespeeds along 
the drift \cite{ftn4}. The resulting mode structure is summarised in 
Fig.\ref{fevol} 
and, from (\ref{xeqn}) and (\ref{yeqn}), in the small-wavenumber 
dispersion relation 
\begin{eqnarray}
\label{disp1}
2 \omega = -(v_1 + v_4)k \sin \theta \pm v_o k -i(2D_T + D_L)k^2 
\nonumber \\
\pm i k^2 D_L \sin \theta \left[{{v_1 - v_4} \over v_o} 
\cos 2 \theta + 2{{v_2 + v_3} \over v_o}  
\cos^2 \theta \right] 
\end{eqnarray}
between frequency $\omega$ and wavenumber $k$, 
where 
\begin{equation}
\label{disp2}
v_o = \sqrt{(v_1 - v_4)^2 \sin^2 \theta + 4 v_2 v_3 \cos^2 \theta},  
\end{equation}
and $\theta$ is the circular polar angle.  
We estimate the resulting wave-speeds 
for NbSe$_2$ in the mixed phase \cite{sb}. Its TDGL
parameters are: $\lambda_H \sim 700$ \AA, 
$\xi \sim 80 \AA$, $T_c \approx 7\mbox{ K }$, $\rho^{(n)}=5\,
\mu\Omega\mbox{ cm}$. For an applied transport
current ${\bf J}_t=1 \mbox{ A}/\mbox{cm}^2$ and inter-vortex separation 
$a\sim \lambda_H$, the wave-speeds $c_{\pm} \sim 1\mu \mbox{m}/\mbox{sec}$.

The most obvious physical consequence of these waves is that 
the dynamic structure factor of a drifting flux lattice 
should display peaks at nonzero frequency (see (\ref{disp1})).  
More dramatically,  
if a region of the flux-lattice moves past an impurity site, 
the impurity will ``pluck'' the flux lattice, and the effect 
will propagate {\em along} and {\em transverse} to the axis of drift,  
shaking up the lattice globally, through  
the sequence of events depicted in Fig.2. 
This wave propagation in the absence of inertia is remarkable,   
and could well be a mechanism for nonthermal noise in drifting 
flux lattices. In addition, time-dependent external disturbances could 
excite resonances with the wavelike normal modes. 

Let us estimate the length scale $\ell_c$ above which these modes are 
actually propagative in character. For wavevectors ${\bf k}
=(k_x,0)$ we see that  
\begin{equation}
\label{lc}
\ell_c \sim \frac{\pi D_L}{\sqrt{v_2 v_3}}
\end{equation}
$ D_L \sim M \lambda$ [see after (\ref{xeqn}), (\ref{yeqn})] 
and $v_i \sim M F$ where $F = J_t \phi_0 /c$ is the Lorentz force per 
unit length on a vortex. Then  
\begin{equation}
\label{lca}
{\ell_c \over a} \sim \frac{\lambda}{F}.  
\end{equation}
$\lambda \approx a H^2/ 8 \pi $ \cite{brandt}, 
$a$ being the flux-lattice spacing, so for $a \sim 10 \mu m$ and 
applied currents $J_t \sim 100$ A/cm$^2$, $\ell_c/a \sim 1$, and  
the propagating modes should dominate.  
However, if $a \sim 10^3 \AA$, $\ell_c / a \sim 10^6$.  
     
In closing, we remark that our work settles an important issue
 in the {\em theory} of the dynamics of moving flux lattices, 
namely their stability \cite{yaron}. We have shown that dynamic 
interactions 
between vortices in a drifting flux lattice without inertia or pinning 
lead to a steady state with {\em stable} linear-response properties. 
Small disturbances about the drifting state travel as {\em waves} with 
a direction-dependent speed which, when calculated in terms of the 
parameters in the TDGL equations, turns out to be a few $\mu$m/s. 
These waves should be observable in systems with large flux-lattice spacing, at large imposed transport currents. The fast scanning tunneling
microscopy approach of Troyanovskii et al. \cite{fstm} seems to be the ideal 
way to observe these waves directly. 
 
We thank D. Gaitonde, T.V. Ramakrishnan, C. Dasgupta, and S. Bhattacharya 
for useful discussions.

\end{multicols} 

\begin{references}
\bibitem{ftn2}These damping 
terms are correct to zeroth order in the drift speed.
\bibitem{tdgl} we adopt the TDGL equations originally proposed by Schmid
	but with a complex relaxation rate. These equations are known to 
	be valid for Type II superconductors with paramagnetic 
	impurities. See also \cite{schmid,dor}.
\bibitem{ftn3} In a suspension of particles drifting
through a viscous fluid the $\{v_i\}$ are determined 
by the {\em hydrodynamic} interaction, which is 
screened only on the scale of the sample thickness.  
In the present problem, that interaction is 
negligible because the ions are a sink for the
momentum of the charged superfluid.
\bibitem{ftn4}Similar arguments 
can be used to deduce the coefficients of nonlinear terms in (\ref{xeqn})
and (\ref{yeqn}).
\end{references}
\end{document}